*Research Article*

# Robust Method for Semantic Segmentation of Whole-Slide Blood Cell Microscopic Images

**Muhammad Shahzad,[1] Arif Iqbal Umar,[1] Muazzam A. Khan,[2] Syed Hamad Shirazi [![ORCID]],[1] Zakir Khan,[1] and Waqas Yousaf[1]**

[1]*Department of Information Technology, Hazara University Mansehra, Dhodial, Pakistan*
[2]*Department of Computing (SEECS), National University of Sciences & Technology (NUST), Islamabad, Pakistan*

Correspondence should be addressed to Syed Hamad Shirazi; syedhamad@hu.edu.pk

Previous works on segmentation of SEM (scanning electron microscope) blood cell image ignore the semantic segmentation approach of whole-slide blood cell segmentation. In the proposed work, we address the problem of whole-slide blood cell segmentation using the semantic segmentation approach. We design a novel convolutional encoder-decoder framework along with VGG-16 as the pixel-level feature extraction model. The proposed framework comprises 3 main steps: First, all the original images along with manually generated ground truth masks of each blood cell type are passed through the preprocessing stage. In the preprocessing stage, pixel-level labeling, RGB to grayscale conversion of masked image and pixel fusing, and unity mask generation are performed. After that, VGG16 is loaded into the system, which acts as a pretrained pixel-level feature extraction model. In the third step, the training process is initiated on the proposed model. We have evaluated our network performance on three evaluation metrics. We obtained outstanding results with respect to classwise, as well as global and mean accuracies. Our system achieved classwise accuracies of 97.45%, 93.34%, and 85.11% for RBCs, WBCs, and platelets, respectively, while global and mean accuracies remain 97.18% and 91.96%, respectively.

## 1. Introduction

Blood is the most delocalized liquid in the body, delivering oxygenated blood from the respiratory system to the other parts of the body and transporting carbon dioxide back [1]. It also helps in the excretion of wastes through the kidney and carries nutrients from the digestive system to the other tissues of the body [2]. Human blood consists of red blood cells (RBCs, erythrocytes), white blood cells (WBCs, leukocytes), and platelets with the ratio of 40% RBCs and 60% WBCs and platelets. Accurate segmentation and classification of RBCs and WBCs play a vital role in the identification of blood-related diseases like blood cancer, syndrome, leukemia, anemia, AIDS, and malaria. The basic determination about classification and segmentation of blood cells is the correct identification of blood constituents and extraction of useful information from the microscopic image.

The blood analysis can be accomplished either by manual counting methods or machine-based methods. Manual counting methods are very tedious, long, subjected to error, and prone with respect to the hematologist expertise. This makes an automatic blood cell image analysis essential for correct and efficient identification of blood-related abnormalities. Currently, several color space techniques are used for the partitioning of microscopic images like RGB, HIS, and LAB, but RGB is the most commonly used color space technique due to its close association with the human visual system [3]. Most of the previous works target the single cell segmentation [4–7]. The single cell technique targets only one type of cell (WBCs or RBCs) for segmentation at one time. But, none of the previous techniques used the whole-slide segmentation of all three types of blood cells simultaneously. Semantic segmentation became popular for dense pixel prediction of objects in images using fully convolutional networks (FCNs) [8, 9]. These networks are penetrated deeply for the detection, classification, and prediction of the pixel base region of interest (ROI). Convolutional neural networks not only improve the local region of interest [10, 11] but also give enormous progress for whole-image segmentation. Semantic segmentation shows the inheritance behavior [11] among the ROI location and semantics. Semantic segmentation addresses the what (semantics) and where (location) questions in the input image. Semantics and location information are encoded in a nonlinear local-to-global pyramid fashion through deepfeature extraction. We focus on the segmentation of WBCs, RBCs, and platelets within a whole-slide image of the cell using the semantic segmentation technique.

In this work, we are going to introduce a novel algorithm for the semantic segmentation of whole-slide RBC, WBC, and platelet images along with development of the state-of-the-art mask-based manually designed blood cell dataset extended from ALL-IDB [12] as the standard for validation of semantic segmentation. The basic motivation behind semantic segmentation is that it is capable of segmenting an unknown image into different portions or objects. The main contributions of this work are as follows:

(1) A robust algorithm for accurate and efficient segmentation of whole-slide WBC, RBC, and platelet cells based on semantic segmentation.

(2) Development of the state-of-the-art manually generated mask-based blood cell dataset extended from ALL-IDB. This dataset includes individual mask of each type of blood cells, i.e., WBCs, RBCs, and platelets, along with whole-slide image mask. Pixelwise labeling is performed on each type of blood cell mask.

(3) A technique that predicts the pixel base ratio of WBCs, RBCs, and platelets in the whole-slide image. This technique plays an important role in the accurate and efficient blood counting during the diagnosis of different types of diseases.

The rest of the paper is organized as follows: Section 2 describes the state-of-the-art previous work, Section 3 describes the approach used for the current work, while data collection/preparation, experimental work, results, and conclusion are described in Sections 4–7, respectively.

## 2. Related Work

Previous work on blood cells like white blood cells (WBCs) and red blood cells (RBCs) [4–7] has used K-means, Zack algorithm, gradient magnitude, watershed transform, and SVM for segmentation along with some preprocessing for image enhancement [13]. These works show outstanding performance for efficient detection and segmentation of blood cells. But most of the work discusses the treatment of single cell from the image (WBCs or RBCs). In [14, 15], Quiñones et al. and Shahin et al. developed an algorithm for the counting and segmentation of leukocytes (WBCs) by using HSV color space/Zack algorithm and adaptive neutrosophic similarity score/Otsu's thresholding, respectively. In [15], BS_DB3 and ALL_DB1 and ALL_DB2 [12] datasets were used for segmentation purposes, while Quiñoneset al. [14] used a total of 12 blood smear images for counting specific type of blood cells in an image. Liu et al. [16] performed segmentation of white blood cells using mean shift clustering and watershed operation on 306 images collected from Hospital of Shandong University. All of them performed analysis on single type of blood cells. No one discussed the segmentation of whole-slide blood cell image into WBCs, RBCs, and platelets. Miao and Xiao [5] performed segmentation on the cropped cells of WBCs and RBCs simultaneously not on whole-slide image with 97.2% and 94.8% accuracy, but they ignored the platelets and also performed negatively on low-quality images. Their algorithm shows that the undersegmentation and oversegmentation rates of RBCs are 1% and 3%, respectively, which are quite high only in 100 images of dataset. Chen et al. [17] developed a framework for the synergistic image and feature adoption based on cross-modality adoption for the segmentation of CT and MR images. The proposed model recovers the performance degradation between 17.2% and 73.0%. Liu et al. [16] performed segmentation of white blood cells using mean shift clustering and watershed operation on 306 images collected from Hospital of Shandong University. Shirazi et al. [6] proposed a hybrid technique by combining the snake algorithm and Gram–Schmidt orthogonalization [18] for the segmentation of WBCs. They performed preprocessing on the input image by the curvelet transform through the Wiener filter to enhance the image for better results. All the analysis was based on single cell. Cao et al. [19] used 203 manually drawn images and 70853 images from Zhongnan Hospital of Wuhan University for the segmentation of leukocytes. They used combination of SWAM&IVFS and fuzzy divergence-based algorithms and got 93.75% segmentation accuracy of WBCs only. White blood cell counting was performed in [20] on ALL-IDB1, 2 datasets using SVM [21] and NNS (nearest neighbor search) with Euclidean distance. This technique only outperforms on WBCs not for RBCs and platelets. In [14, 15], Quiñones et al. and Shahin et al. developed an algorithm for the counting and segmentation of leukocytes (WBCs) by using HSV color space/Zack algorithm and adaptive neutrosophic similarity score/Otsu's thresholding, respectively. Yi et al. [22] segmented only RBCs with the help of the markercontrolled watershed transform algorithm on manually collected 117 images. They select area, perimeter, and circulatory parameter for the identification of red blood cells. Image enhancement is also carried out as the preprocessing step with watershed transform algorithm [23]. Normoblast cells were identified automatically by Das et al. [24] by performing experiments on the 950 nucleated blood cells. The marker-controlled watershed algorithm was used for segmentation of RBCs. Mean intensity, standard deviation, skewness, kurtosis, and entropy features were extracted for correct and efficient RBC identification. Shirazi et al. [25] segmented the RBCs using the statistical-based thresholding method/fuzzy c-means on single blood cell on the ALL-IDB dataset. Texture and geometrical features were extracted for separation of targeted cells. Area, perimeter, circulatory, convexity, and solidity parameters were extracted in [26] for segmentation of WBCs (lymphocytes, neutrophils, eosinophils, and basophils) from 117 images collected by the authors.

## 3. Dataset Preparation

We use the acute lymphoblastic leukemia image database (ALL-IDB1) [12] as the baseline dataset for our experimental work. It consists of a total of 108 whole-slide blood cell images. 108 images contain about 39000 blood elements. All the images were taken with 300 to 500 magnification rate microscopes, out of which 59 (2592 × 1944) cells were from healthy individuals and 49 (1712 × 1368) from acute lymphoblastic leukemia (ALL) patients.

In Figure 1, images from Figures 1(a)–1(c) are taken from healthy individuals, while images from Figures 1(d)–1(f) are taken from acute lymphoblast leukemia patients.

In order to perform semantic segmentation on these images, we extend this dataset by generating the state-of-the-art fine tune mask image of each of 39000 blood elements from 108 SEM images. We generate a total of 432 masks (Figures 2–5). 108 individual masks of each blood cell type (WBCs, RBCs, and platelets) along with 108 combine masks without background. Pixels of each mask image are labeled according to their blood cell type. All these masks work as ground truth images during training and testing process. All the extended pixel-level mask images are available at LINK.

In Figure 2, the images shown in Figure 2(a) are taken from ALL-IDB. Figure 2(b) represents the RBC individual mask (ground truth) from the original image. Rest of the cells are ignored. However, Figure 2(c) represents pixelwise labeling of each RBC. All the pixels where RBCs are present are set as black, otherwise white.

In Figure 3, images shown in Figure 3(a) are taken from ALL-IDB. Figure 3(b) represents the WBC individual mask (ground truth) from the original image. Rest of the cells are ignored. However, Figure 3(c) represents pixelwise labeling of each WBC cell. All the pixels where WBCs are present are set as black, otherwise white.

In Figure 4, images shown in Figure 4(a) are taken from ALL-IDB. Figure 4(b) represents the platelet individual mask (ground truth) from the original image. Rest of the cells are ignored. However, Figure 4(c) represents pixelwise labeling of each platelet. All the pixels where platelets are present are set as black, otherwise white.

In Figure 5, images shown in Figure 5(a) are taken from ALL-IDB. Figure 5(b) represents the combine mask (ground truth) of each type of cell from the original image. Rest of the area that represents background is ignored. However, Figure 5(c) represents pixelwise labeling of each type of cell. In the combined mask, each type of cell is given a specific ID for representation during pixel base labeling, i.e.,

```
Masking value replacement with respective IDs
MaskAll (find (RBC_M > 0)) = 1 (Red)
MaskAll (find (WBC_M > 0)) = 2 (Blue)
```

```
MaskAll (find (PLT_M > 0)) = 3 (Green)
```

## 4. Methodology/Our Approach

*4.1. Preprocessing.* To prepare the input images according to the standard input of the proposed system, preprocessing is applied on the dataset.

Preprocessing comprises 8 steps shown in Figure 6: (1) System reads original and fine-tuned manually generated mask images from memory. (2) All the masked images are checked for 3 channels of color. If any image found RGB, it will convert into 2 channel images for further processing with original image. Fusing of mask pixel labeling of each type is also performed in this step. (3) The complement of ROI is found, value 0 is assigned to the rest of the pixels, and unity mask is generated. (4) The segmented image is generated with all pixel values zero. (5) In this step, we find the pixels of RBCs, WBCs, and platelet individually. (6) All pixels related to the RBCs are assigned with pixel ID, PID = 1, WBCs with PID = 2, and platelets with PID = 3. (7) In this step, the pixel-labeled images are written on the memory. (8) In this step, resizing of original and masked images is done according to the framework requirement and written on the memory.

*4.2. Semantic Segmentation.* Semantic segmentation is penetrated deeply for the detection, classification, and prediction of pixel-based region of interest (ROI) in an image. For accurate and proper segmentation of whole-slide blood cell, we design semantic segmentation-based framework along with VGG16 [10] as pretrained feature extraction model. Figure 7 shows the main architecture and work flow of the proposed framework.

*4.3. Pseudocode of Semantic Segmentation*

(1) START

(2) Load dataset (original and masked)

   Dataset ⟵ All IDB1

   Read Images (x)

   Count Labels y ⟵ 3

(3) Preprocessing

   (a) Fusing mask pixel labeling

      (i) Pixels complement
      (ii) Unity mask generation

   (b) Segmented images generation with zero-pixel values

      (i) Find segmented areas of WBCs, RBCs, and platelets
      (ii) Assign pixel IDs (RBC = 1, WBC = 2, and platelet = 3)

   (c) Resize original image and masked images into 300 × 300 × 3

   (d) Write images and masks to disk as datastore

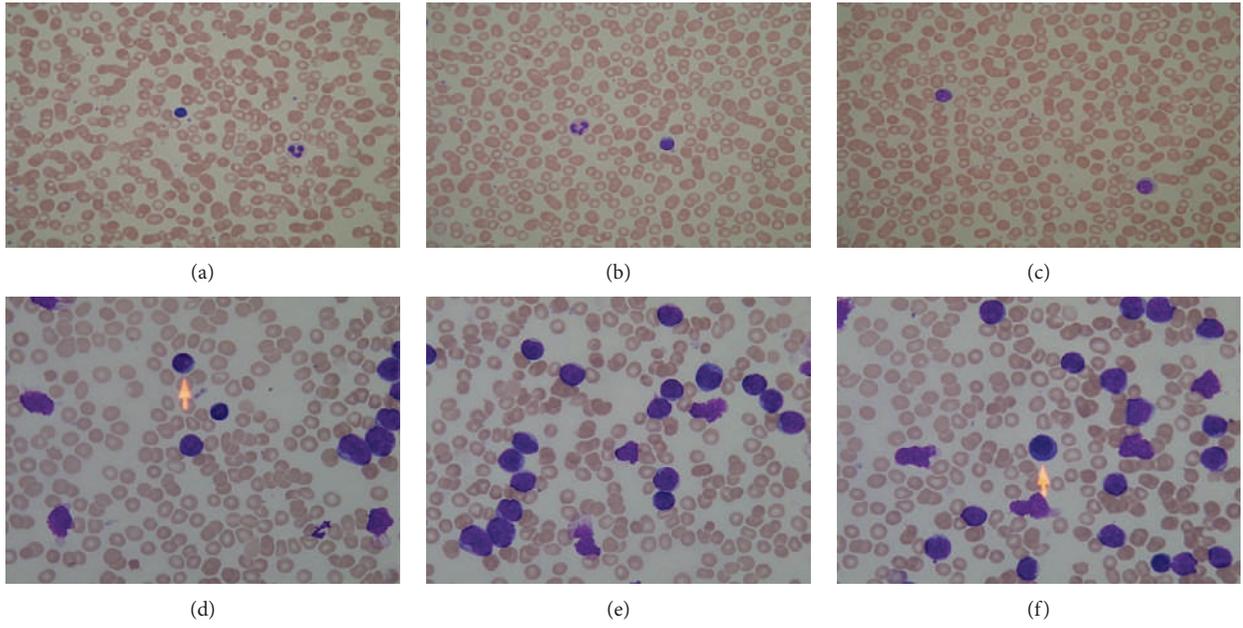

Figure 1: Sample images from ALL-IDB1 datasets.

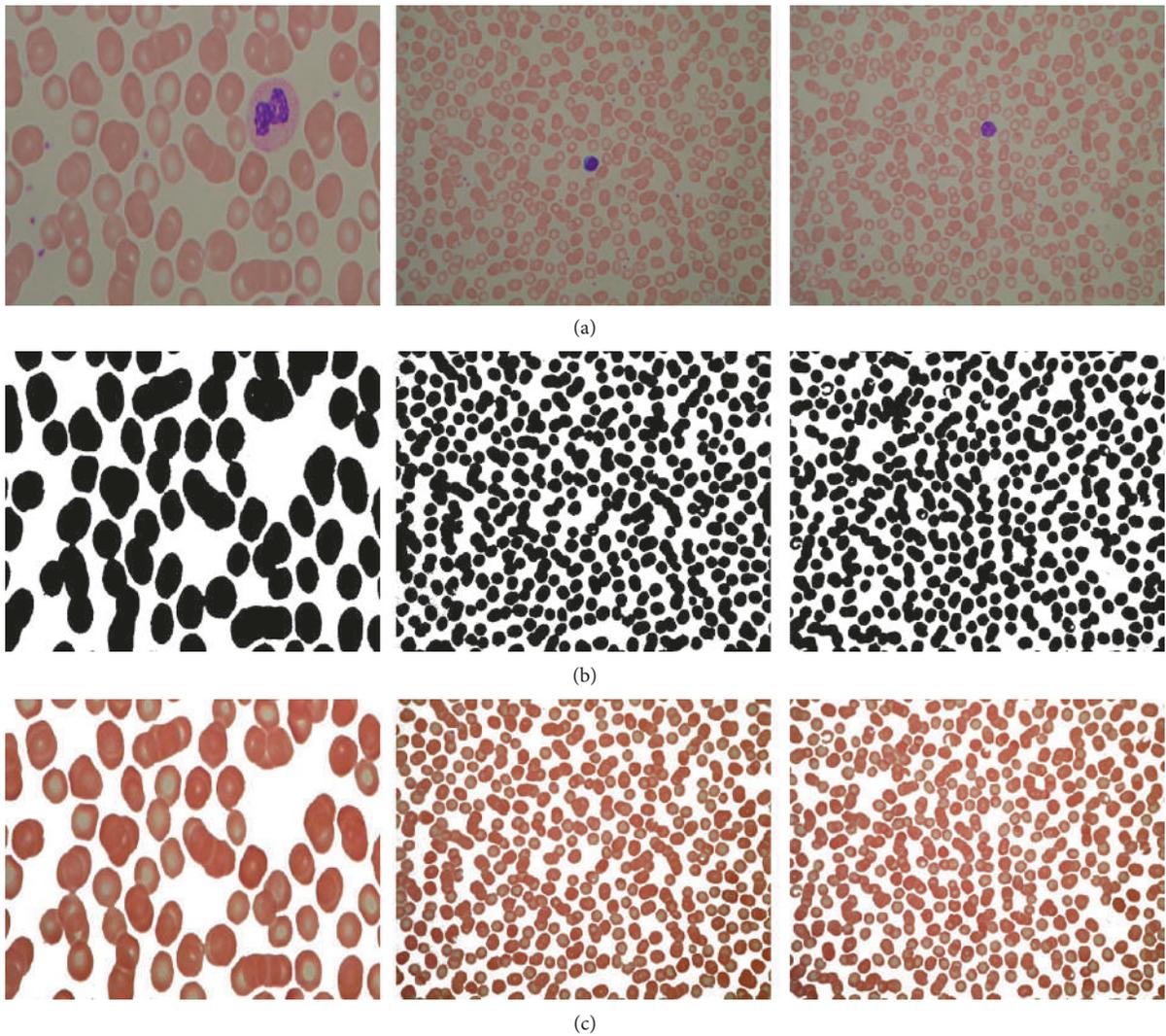

Figure 2: RBC masks and pixelwise labeling. (a) Original image. (b) Platelet cell mask. (c) Pixelwise labeling.

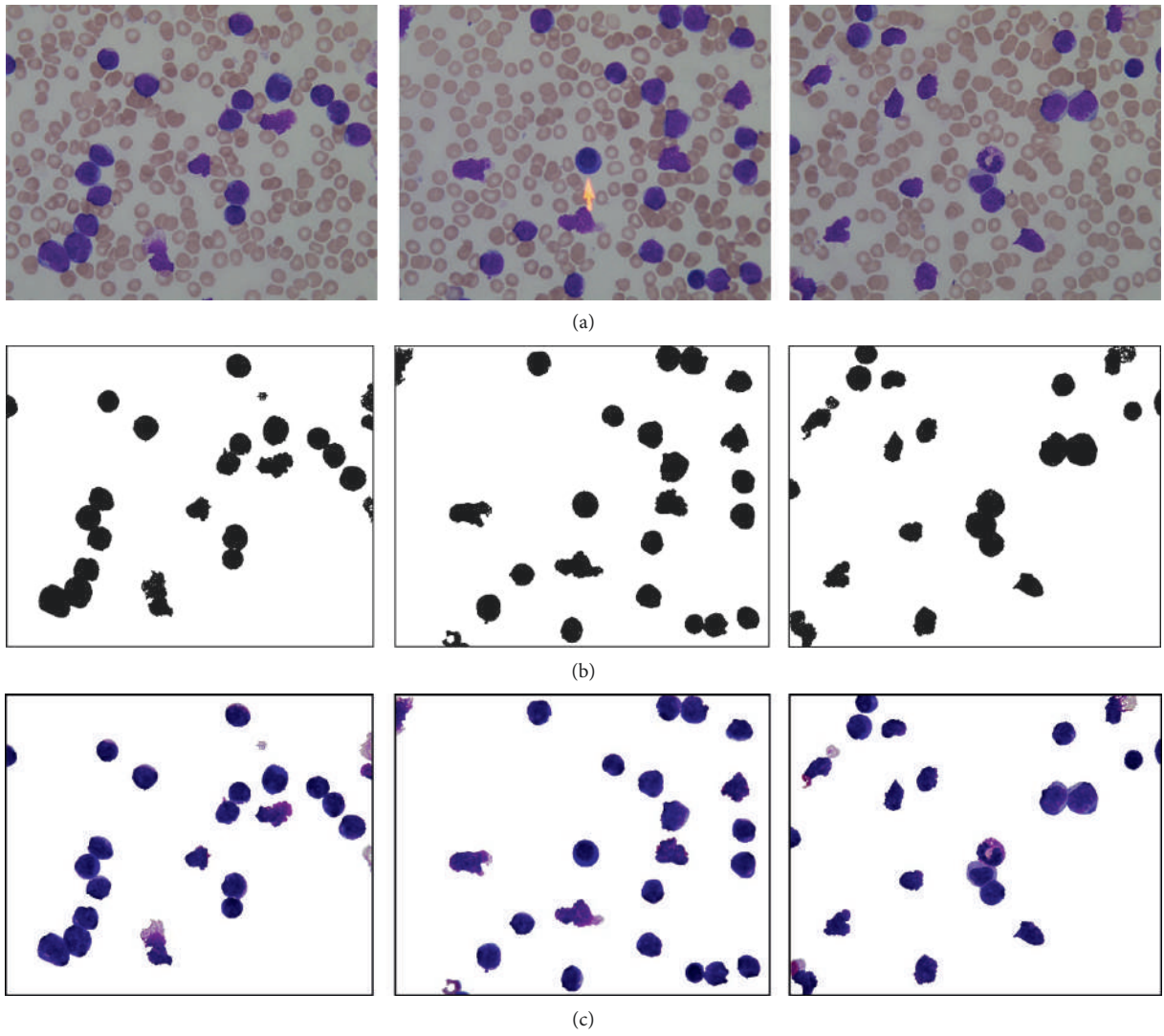

Figure 3: WBC masks and pixelwise labeling. (a) Original image. (b) Platelet cell mask. (c) Pixelwise labeling.

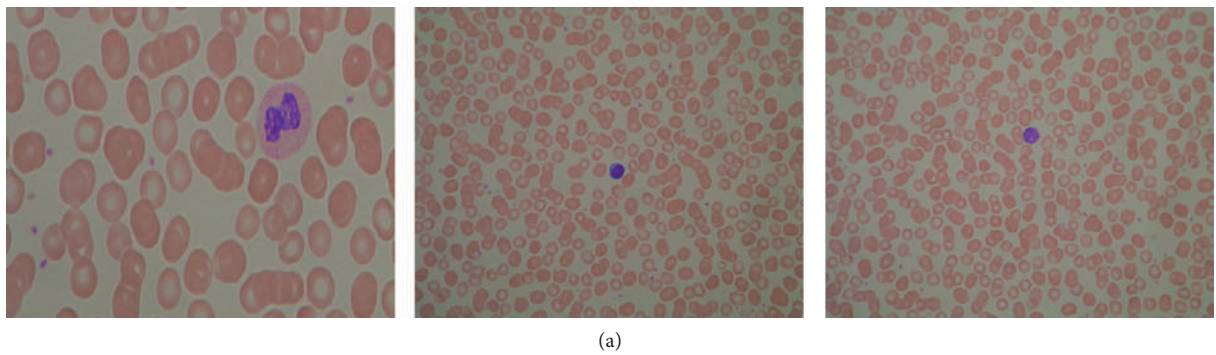

(a)

Figure 4: Continued.

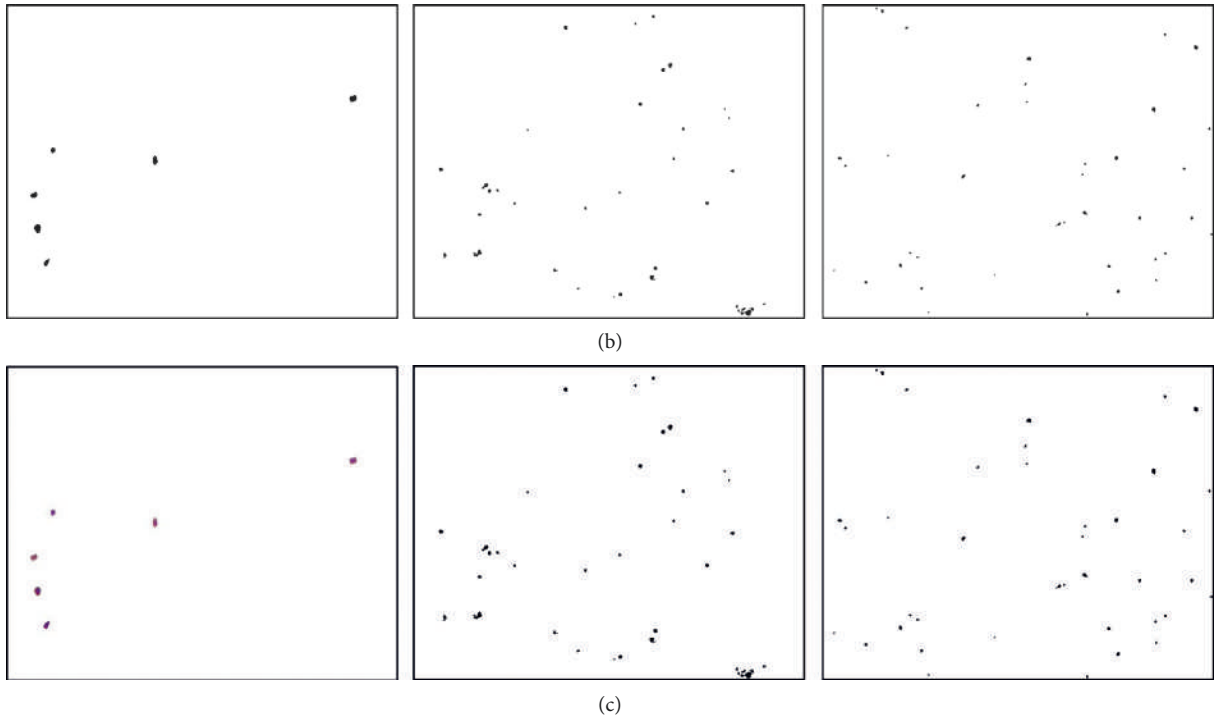

Figure 4: Platelet masks and pixelwise labeling. (a) Original image. (b) Platelet cell mask. (c) Pixelwise labeling.

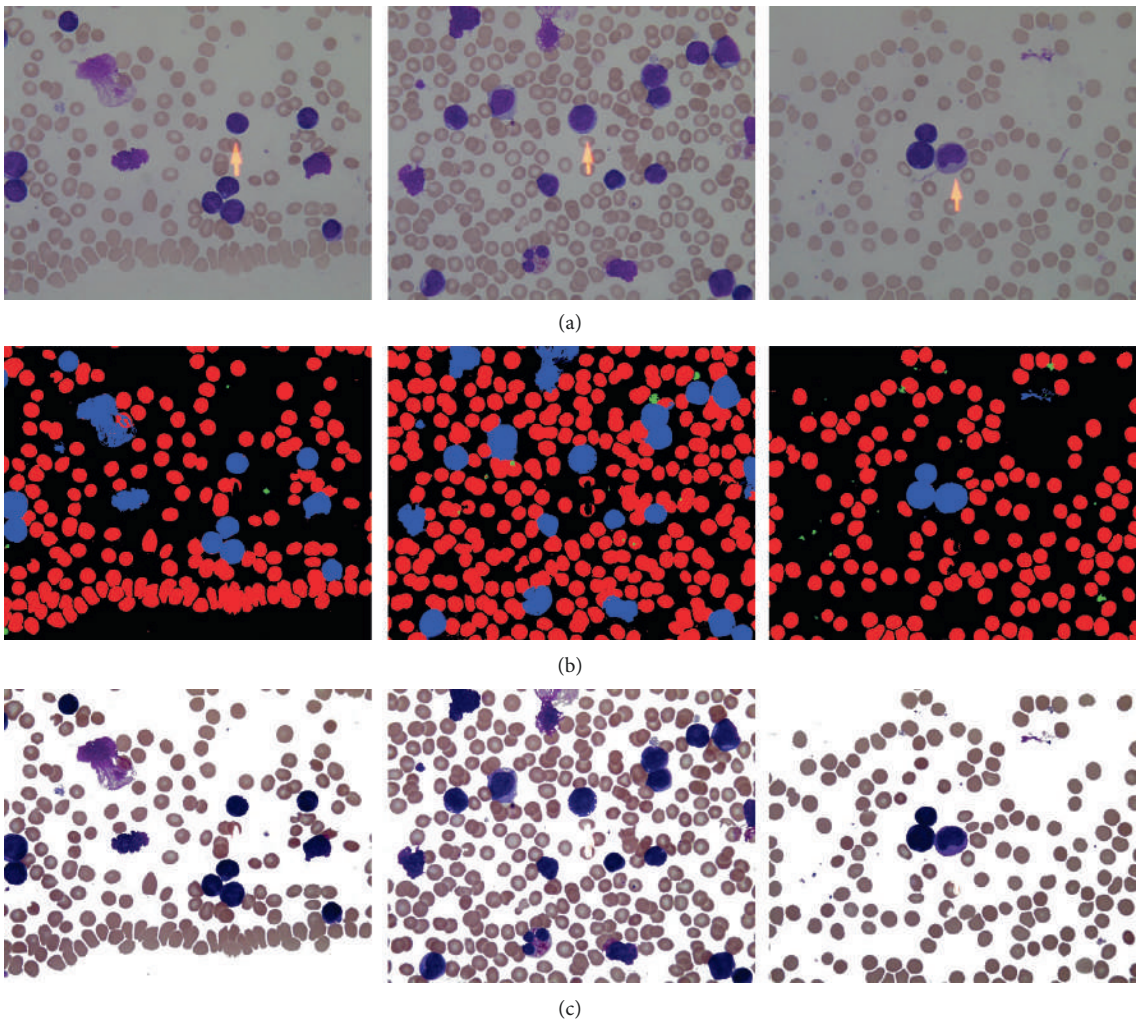

Figure 5: Whole-slide image masks and pixelwise labeling. (a) Original image. (b) Platelet cell mask. (c) Pixelwise labeling.

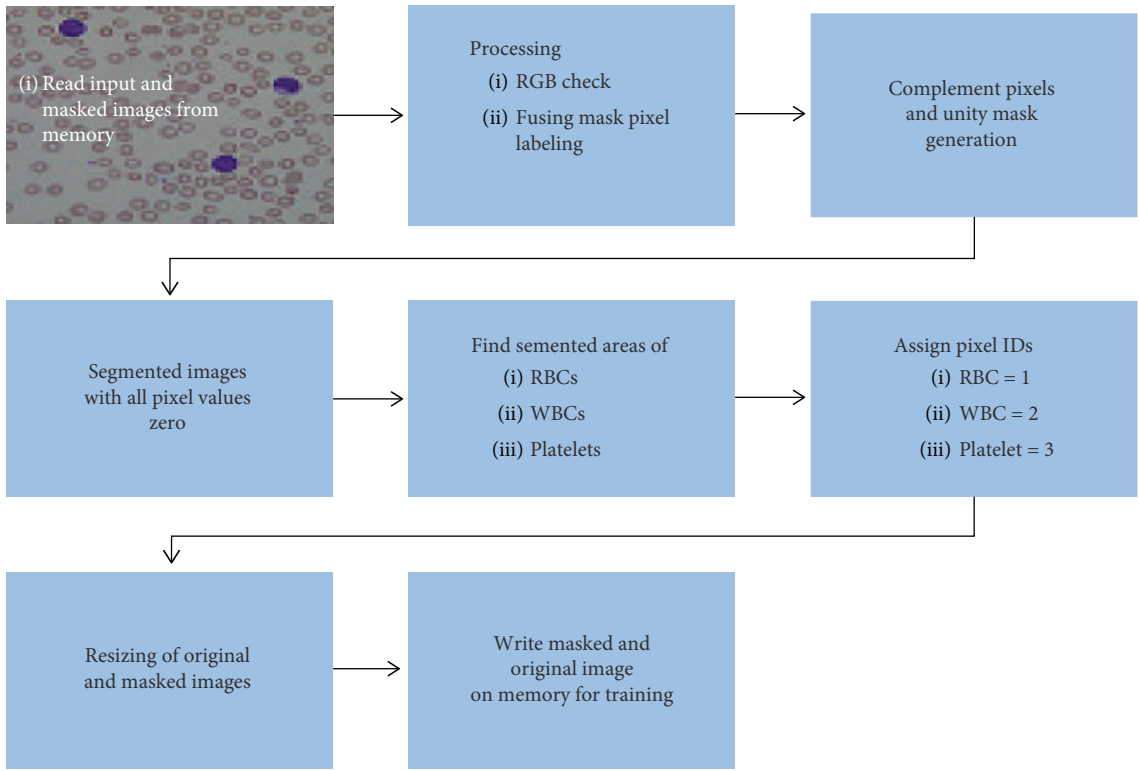

FIGURE 6: Flow chart of preprocessing.

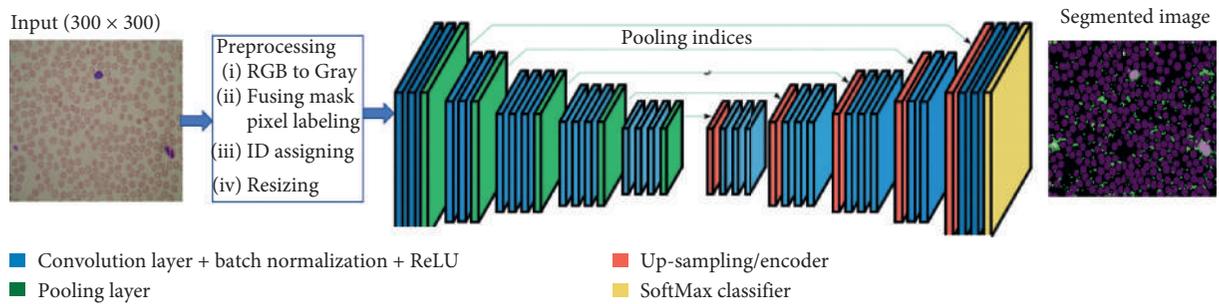

FIGURE 7: A deep convolutional encoder-decoder (DCED) network for semantic segmentation.

(4) Load pretrained VGG16 for pixel-level feature extraction

(5) Load blood cells datastore

  (a) Original images
  (b) Ground truth

(6) Split datastore into train and test sets

(7) Load the proposed DCED network

(8) Training options

  (i) Min_Batch_Size ⟵ 1
  (ii) Epoch ⟵ 500
  (iii) Iterations per epoch ⟵ 103
  (iv) Initial learning rate ⟵ $1e-3$

(9) Train the model

(10) Class-wise pixel counting of blood cells

(11) Compute evaluation metrics

  (a) Accuracy ⟵ (TP + TN)/(TP + TN + FP + FN)
  (b) IoU ⟵ Target ∩ Prediction/Target ∪ Prediction
  (c) BF Score ⟵ 2(Precision ∗ Recall)/(Precision + Recall)

(12) END

Our semantic segmentation technique comprises 5 steps: (1) Initially, original images along with manually generated mask are uploaded to the system for preprocessing. (2) During preprocessing, the first step is the fusing of mask pixel labels for pixel complement and unity mask generation. The next step is the generation of the segmented image with all pixel values zero. Then, we find the separate segmented areas of each type of blood cell using manually segmented images and assign specific pixel ID to each cell's type, i.e., RBCs = 1, WBCs = 2, and platelets = 3. (3) After that, each original image and masked labeled images are resized according to the standard network input. (4) In his step, VGG16 is loaded as the pretrained feature extraction model. (5) Blood cell is split into train and test images and loaded in the datastore. (6) Now the proposed framework DCED network is loaded for training. (7) After training process, segmented whole-slide blood images are produced along with pixel level counting of each type of blood cell. For validation purposes, three types of evaluation matrices are used during the testing phase. These matrices are accuracy, IoU, and BF score. We perform semantic segmentation on the whole-slide blood cell image. We also propose a novel and state-of-the-art technique for blood cell type counting on pixel level that is very helpful for disease diagnosis related to blood cells and cell counting. This technique gives exact number of blood cell counting.

## 5. Training and Evaluation Matrices

*5.1. Training.* Our proposed model is trained and tested on the ALL-IDB1 dataset. A total of 108 whole-slide blood cells with 39000 blood elements are used, out of which the system chooses randomly 103 images for training and 5 images for testing purposes. We train our model on 500 epochs along with 103 iterations in each epoch and minibatch size of one. The initial learning rate is set as $1e^{-3}$. The segmented image is obtained at the last layer of the decoding part of architecture. The system is trained on Windows 10 operating system with 24 GB RAM and 2 GB NVIDIA 750Ti single GPU.

*5.2. Evaluation.* This section describes all the results obtained during the training and testing phases.

*5.2.1. Epochs vs Accuracy and Epochs vs Loss Projection.* In Figure 8, Figure 8(a) shows the ratio between the number of epochs used during training and accuracy achieved against each epoch. Figure 8(b) shows the ratio between the number of epochs and loss against each epoch.

Figures 6 and 7 illustrate the fine tuning of the proposed model (DCED framework) during 500 epochs on the dataset. The graph shows that accuracy and corresponding loss of training process started from 41.38% and 1.21%, respectively, at first epoch. But, after training of 10 epochs, the accuracy goes to 80.56%, while corresponding loss decreases up to 0.65. After that, the accuracy fluctuates between 83% and 92.56%. Highest accuracy is achieved at the epoch number 134, while minimum validation loss is achieved at the epoch number 428 with 0.0230 rate.

*5.2.2. Iteration vs Accuracy and Iteration vs Loss Projection.* We train our model on 500 epochs. Each epoch consists of 103 iterations. The total number of iteration during training was 51500. Following graphs, Figure 9 shows the accuracy and loss against the iteration.

Figure 9(a) shows the ratio between the number of iteration and accuracy achieved against iteration. Figure 9(b) shows the ratio between the number of iteration and loss against each iteration.

At iteration number 1, the accuracy plot started from 0 and directly jumped to 40.18% with 1.2 loss. After 1000 iterations of the training data, accuracy increased up to 80.56% while loss decreased to 0.65. After 14000 iterations, the peak of the graph showed the highest accuracy rate, while after 44000 iterations, the system attained the lowest rate of loss with 0.023 loss. Figure 10 gives a clear picture of both the consistency in accuracy and loss during the training phase.

Figure 11 elaborates the frequency of each blood cell element, i.e., WBCs, RBCs, and platelets, in one image. It contains 93.55% RBCs, 6.09% WBCs, and 0.34% platelets.

*5.2.3. Evaluation Matrices.* We evaluate our model on 3 evaluation matrices: (1) accuracy (global and mean); (2) intersection over union (IoU) (MeanIoU and WeightedIoU); and (3) MeanBFScore.

*5.2.4. Accuracy.* In semantic segmentation, percent of correctly classified pixels are determined by calculating accuracy. It is the ratio between correctly identified positive pixels (TP), correctly identified negative pixels (TN) over TP, TN along with falsely identified positive (FP), and falsely identified negative (FN):

$$\text{accuracy} = \frac{\text{TP} + \text{TN}}{\text{TP} + \text{TN} + \text{FP} + \text{FN}}. \quad (1)$$

We calculate accuracy of each class of the blood cells, i.e., RBCs, WBCs, and platelets, along with global accuracy of all classes. Individual class accuracy is shown in Table 1. We have got accuracies of 97.45%, 93.34%, and 85.11% for RBCs, WBCs, and platelets, respectively. Our algorithm achieved 97.18% global accuracy and 91.96% mean accuracy, which is shown in Table 2.

*5.2.5. Intersection over Union (IoU).* Intersection over union (IoU) [27] is a class of image segmentation evaluation matrix that quantifies the ratio between overlapping of target ground truth mask and prediction output. It is calculated by finding the ratio between the intersection of target and prediction pixels over all pixels in both masks:

$$\text{IoU} = \frac{\text{Target} \cap \text{Prediction}}{\text{Target} \cup \text{Prediction}}. \quad (2)$$

We have got IoU of 0.54431, 0.40626, and 0.009304 for RBCs, WBCs, and platelets, respectively, shown in Table 1, while MeanIoU and WeightedIoU were 0.31996 and 0.53511, respectively.

*5.2.6. BF Score.* It measures the percent of boundary match between ground truth boundary and predicted boundary of an object [28]. It is a combined ratio of twice of precision and recall product over sum of recall and precision:

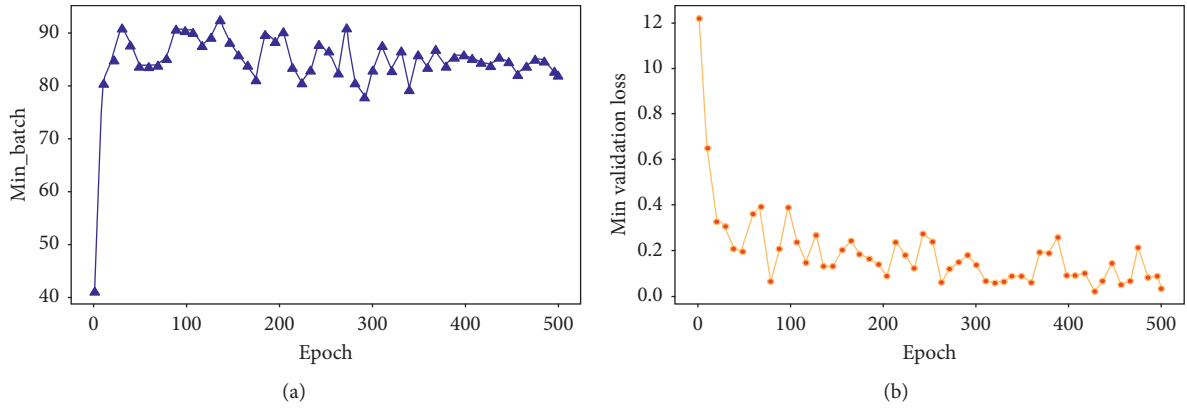

Figure 8: Epoch vs (a) accuracy projection and (b) loss projection.

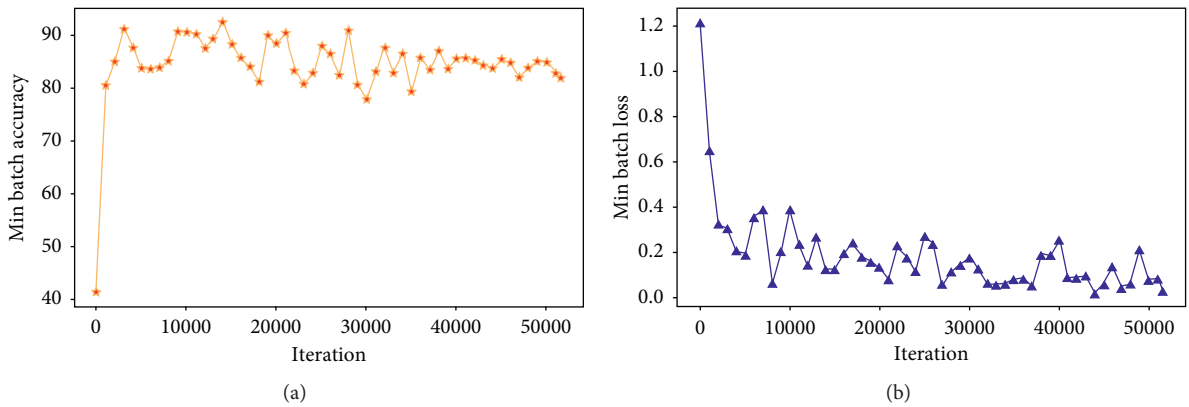

Figure 9: Iteration vs (a) accuracy projection and (b) loss projection.

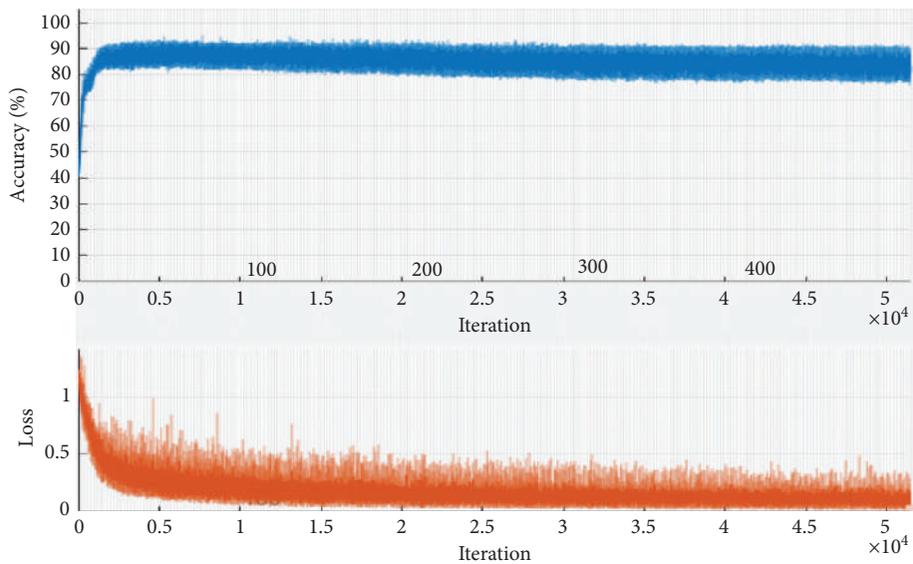

Figure 10: Training progress of the proposed model.

$$BF\ score = \frac{2 * (Precision * Recall)}{Recall + Precision}. \quad (3)$$

Classwise MeanBFScore of our technique was 0.59489, 0.33086, and 0.15307 for RBCs, WBCs, and platelets, respectively, while overall MeanBFScore was 0.40654.

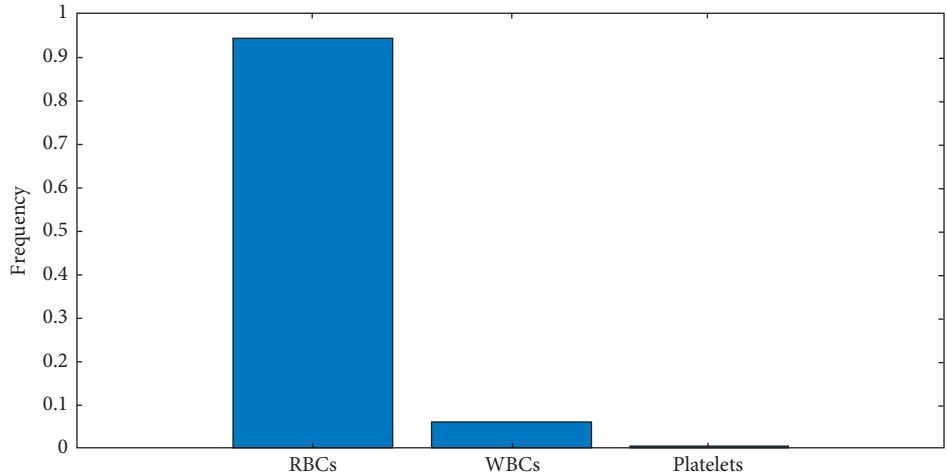

FIGURE 11: Frequency of blood cells.

TABLE 1: Classwise accuracy, intersection of union, and Mean-BFScore of RBCs, WBCs, and platelets.

| S. no. | Class | Accuracy | IoU | MeanBFScore |
|---|---|---|---|---|
| 1 | RBCs | 0.97451 | 0.54431 | 0.59489 |
| 2 | WBCs | 0.93342 | 0.40626 | 0.33086 |
| 3 | Platelets | 0.85112 | 0.009304 | 0.15307 |

TABLE 2: Global accuracy, mean accuracy, MeanIoU, WeightedIoU, and MeanBFScore.

| Evaluation matrix | Value |
|---|---|
| GlobalAccuracy | 0.97184 |
| MeanAccuracy | 0.91969 |
| MeanIoU | 0.31996 |
| WeightedIoU | 0.53511 |
| MeanBFScore | 0.40654 |

## 6. Results and Discussion

The proposed model comprises 2 phases: (1) preprocessing phase, in which original images and ground truth masks are modified according to the required format, and (2) deep convolutional encoder-decoder phase (DCED), which consists of a pretrained VGG16 model for pixel-level feature extraction and encoder-decoder framework for training and testing of blood cells. In order to prove the outperformance of our model, we calculate classwise accuracy, IoU, and MeanBFScore of our model.

### 6.1. Accuracy.
Table 1 shows the classwise accuracy of each blood cell. Our model achieved RBC segmentation accuracy of 97.45%, while WBCs and platelets are 93.34% and 85.11% correctly segmented. Intersection of union (IoU) of the proposed model for RBCs, WBCs, and platelets are 0.54431, 0.40626, and 0.009304, respectively. We got MeanBFScore of RBCs = 0.59489, WBCs = 0.33086, and platelets = 0.15307. Figure 9 shows the accuracy, IoU, and MeanBFScore of each blood cell visually.

Combined analysis of each evaluation matrix is shown in Table 2. Our model attains global and mean accuracies of 0.97184 and 0.91969, respectively. MeanIoU and weighted IoU are 0.31996 and 0.53511, while MeanBFScore of all the three classes is 0.40654 shown in Figure 12.

### 6.2. Classwise Pixel Counting of Blood Cells.
During the blood disease prediction, accurate pixel counts each type of blood cell are very important. Our technique also gives the accurate pixel count of RBCs, WBCs, and platelets. Table 3 shows the number of pixels of each blood element. In Figure 13, the chart shows the number of each blood cell pixel in the test image.

### 6.3. Comparison with Previous Work.
Technique proposed in [19] only targets the RBC segmentation with 93.5% accuracy, while the techniques in [4, 15, 29] target only WBC segmentation with an accuracy rate of 82%, 97.6%, and 90%, respectively. In [5], the author targets RBCs and WBCs with an accuracy rate of 94.8% and 97.2%, respectively. All these segmentations are performed only on single cell. Our model targets the whole-slide image and segments each blood cell type, RBCs, WBCs, and platelets, and gets high accuracy rate (given in bold in Table 4) RBCs = 97.45%, WBCs = 93.34%, and platelets = 85.11%. Our method also outperforms in the context of global accuracy with a percent rate of 97.18%. We also used three types of evaluation matrices to find the effectiveness of the proposed framework. In addition to this, we also introduce a state-of-the-art technique for accurate counting of each blood cell on pixel level that gives important assistance for blood counting tests. Table 4 shows the comparative summary of our model with previous work.

### 6.4. Computational Speed.
The average training time of DCED framework with 500 epochs is 10 hours approximately, while the testing time for one whole-slide blood image is 1.5 second on Windows 10 operating system with 24 GB RAM and 2 GB NVIDIA 750Ti single GPU.

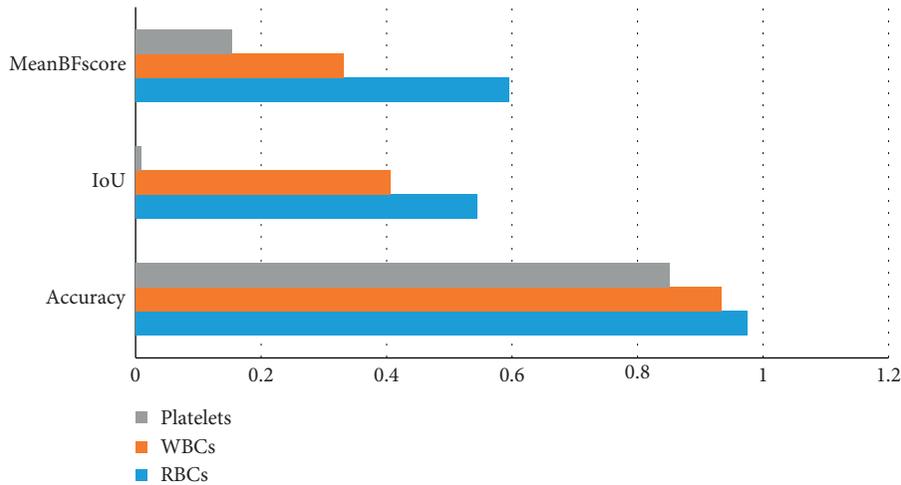

Figure 12: Classwise evaluation matrix charts of RBCs, WBCs, and platelets.

Table 3: Pixels counting of blood cell element.

| S. no. | Name of cell | Pixel count |
|---|---|---|
| 1 | RBCs | $2.3076e + 08$ |
| 2 | WBCs | $1.5036e + 07$ |
| 3 | Platelets | $8.609e + 05$ |

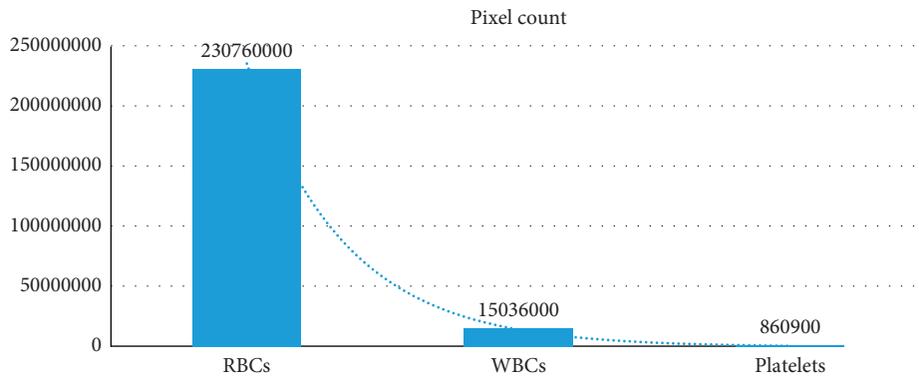

Figure 13: Counting of blood cell element within image.

Table 4: Quantitative comparison of proposed technique with existing techniques.

| Paper | WBCs seg | RBCs seg | Platelets seg | Accuracy (%) | Global accuracy | Evaluation matrices | Cell type |
|---|---|---|---|---|---|---|---|
| Nikitaev et al. [4] | ✓ | ✗ | ✗ | WBCs = 82 | ✗ | Accuracy | Single cell |
| Miao and Xiao [5] | ✓ | ✓ | ✗ | WBCs = 97.2 RBCs = 94.8 | ✗ | Accuracy | Single cell |
| Cao et al. [19] | ✓ | ✗ | ✗ | RBCs = 93.75 | ✗ | Accuracy | Single cell |
| Shahin et al [15] | ✓ | ✗ | ✗ | WBCs = 97.6 | ✗ | Accuracy | Single cell |
| Mohammed et al. [29] | ✓ | ✗ | ✗ | WBCs = 90 | ✗ | Accuracy | Single cell |
| **Proposed method** | ✓ | ✓ | ✓ | **RBCs = 97.45 WBCs = 93.34 Platelets = 85.11** | **97.18%** | **Accuracy, IoU, mean and weighted IoU, BFScore** | **Whole-slide image, segment WBCs, RBCs, and platelets simultaneously** |

*6.5. Advantages over Existing Techniques.* The proposed framework has upper hand on existing work related to blood cell segmentation in respect of the following:

(i) We used whole-slide image segmentation rather than single cell

(ii) We used pixel-level semantic segmentation approach of blood cell rather than object-level classic segmentation methods

(iii) We also proposed state-of-the-art fine-tuned masks of whole-slide blood images

(iv) A novel technique for blood cell counting at pixel level

(v) We used three different evaluation matrices for validation of technique rather than only accuracy

(vi) Less execution time for testing of whole-slide blood cell, i.e., 1 sec per image

The main limitation of this work is that in some cases, it may require huge amount of labeled data like in scene classification. In such type of problems, millions of labelled images are required which are not available.

## 7. Conclusion

This work addresses the problem of semantic segmentation in medical imaging especially in blood cells. To the best of our knowledge, we are the first who target the semantic segmentation of the whole-slide blood cell. The proposed framework is designed for accurate semantic segmentation of the whole-slide blood cells. We proposed a novel convolutional encoder-decoder framework along with VGG16 as pixel-level feature extraction model. Prior to the training process, dataset passed through the preprocessing step where all the original images along with manually generated masks processed for conversion into the format that is suitable for underlying framework. Our system is evaluated on the basis of 03 evaluation metrics. We got outstanding results with respect to class wise as well as global and mean accuracies. Our system achieved classwise accuracies of 97.45%, 93.34%, and 85.11% for RBCs, WBCs, and platelets, respectively, while global and mean accuracy remain 97.18% and 91.96%, respectively. In future, we will perform experiment on dilated and blurred images.

## Data Availability

The xxx.jpg data used to support the findings of this study have been deposited in the Google Drive repository (https://drive.google.com/drive/folders/1F7kZ1SRWUD9R6aHLMkj3wsjcHnvlGuwP?usp=sharing).

## Conflicts of Interest

The authors declare that they have no conflicts of interest.

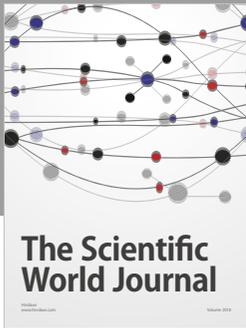
**The Scientific World Journal**

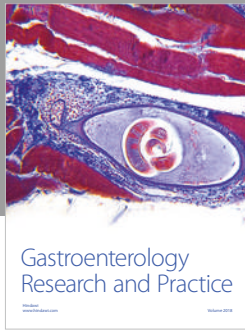
Gastroenterology Research and Practice

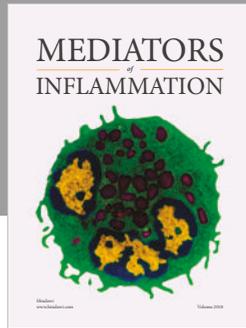
MEDIATORS of INFLAMMATION

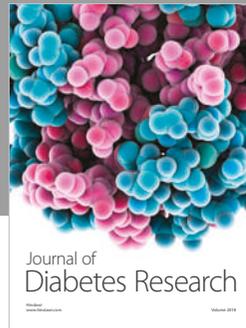
Journal of Diabetes Research

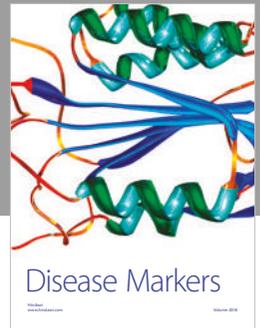
Disease Markers

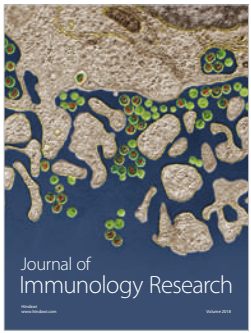
Journal of Immunology Research

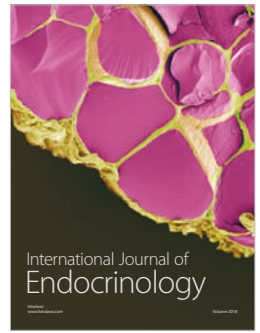
International Journal of Endocrinology

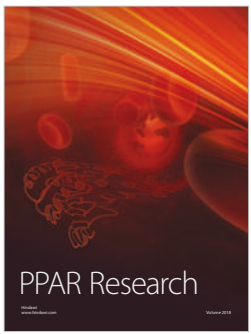
PPAR Research

Submit your manuscripts at
www.hindawi.com

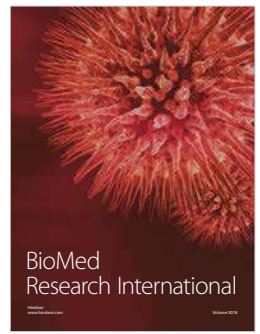
BioMed Research International

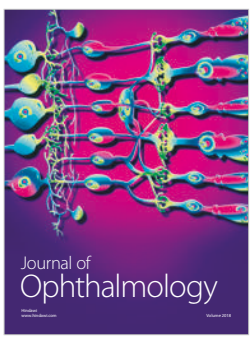
Journal of Ophthalmology

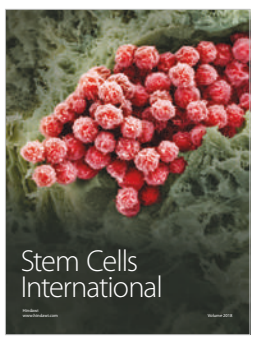
Stem Cells International

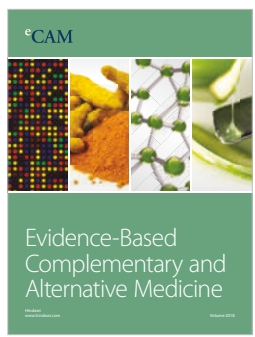
CAM
Evidence-Based Complementary and Alternative Medicine

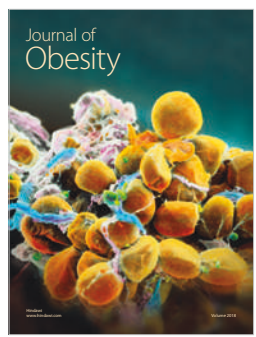
Journal of Obesity

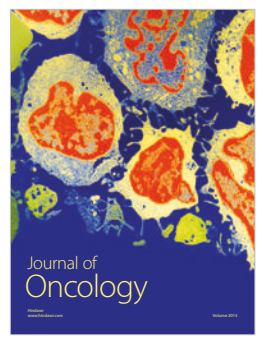
Journal of Oncology

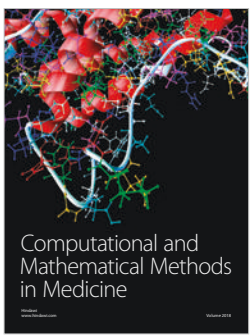
Computational and Mathematical Methods in Medicine

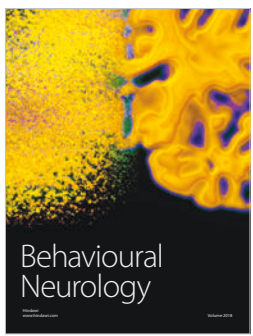
Behavioural Neurology

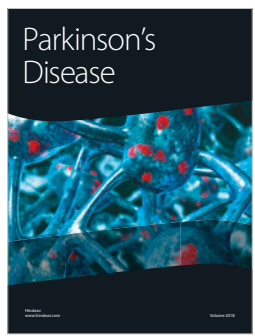
Parkinson's Disease

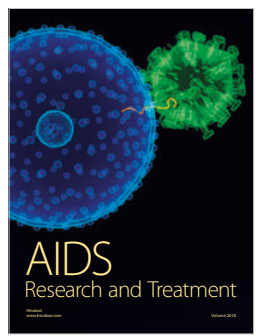
AIDS
Research and Treatment

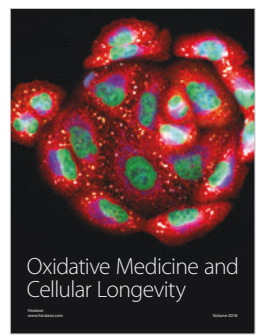
Oxidative Medicine and Cellular Longevity